\documentclass[journal,article,submit,pdftex,moreauthors]{Definitions/mdpi} 

\firstpage{1} 
\pubvolume{1}
\issuenum{1}
\articlenumber{0}
\pubyear{2023}
\copyrightyear{2023}
\datereceived{ } 
\daterevised{ } 
\dateaccepted{ } 
\datepublished{ } 
\hreflink{https://doi.org/} 

\Title{Charm production and hadronisation in pp and p--Pb collisions at $\sqrt{s_{\rm NN}} = $ 5.02 TeV at the LHC with ALICE }

\TitleCitation{Title}

\Author{Tiantian Cheng $^{1,2}$\orcidA{}*, on behalf of the ALICE Collaboration}

\AuthorNames{Tiantian Cheng}

\AuthorCitation{Cheng, T.}

\address{%
$^{1}$ \quad Central China  Normal  University (CCNU), Wuhan, China \\
$^{2}$ \quad  GSI  Helmholtzzentrum  für  Schwerionenforschung (GSI), Darmstadt, Germany}

\corres{Correspondence: tiantian.cheng@cern.ch}

\abstract{Production measurements of charm hadrons, particularly the yield ratios of different hadron species as a function of the transverse momentum, are important to study the charm hadronisation mechanism. In this contribution, measurements of $\rm D^{0}$, $\rm D^{+}$, and $\rm D^{+}_{\rm s}$ meson production, as well as $\Lambda^{+}_{\rm c}$-baryon production at midrapidity in pp and p--Pb collisions at $\sqrt{s_{\rm NN}} = $ 5.02 TeV with ALICE will be discussed. The measurements will be compared with predictions from Monte Carlo event generators and theoretical calculations. In addition, the first baryon-to-meson yield ratio measurements for $\Lambda^{+}_{\rm c}/\rm D^{0}$ down to $p_{\rm T}$ = 0 in p--Pb collisions will be discussed. In such collisions, a modification of the hadronisation mechanisms could be present due to cold nuclear matter effects and possible collective phenomena.  Finally, measurements of charm fragmentation fractions and the total charm production cross section per unit of rapidity at midrapidity, in pp and p--Pb collisions at $\sqrt{s_{\rm NN}} = 5.02$ TeV at the LHC, will be presented.}


\begin{document}
\section{Introduction}

Measurements of the production of heavy-flavour hadrons in high-energy hadronic collisions provide important tests for calculations based on perturbative quantum chromodynamics (pQCD). The production cross section of charm hadrons in hadronic collisions is usually calculated using the factorisation approach as a convolution of three factors: the Parton Distribution Functions (PDFs), the hard-scattering cross section at the partonic level, and the fragmentation functions of the produced heavy quarks into given species of heavy-flavour hadrons, which are assumed to be universal across different collision systems. However, recent observations of enhanced baryon-to-meson production yield ratios in hadronic collisions at low and intermediate $p_{\rm T}$, with respect to the same measurements performed in $\rm e^{+}e^{-}$ or $\rm e^{-}$p collisions, suggest that the charm fragmentation fractions are not universal and depend on the collision system. The study of charm-baryon production in p--Pb collisions can be used to examine possible modifications of their production due to the presence of cold nuclear matter (CNM) effects. Moreover, the  measurement of the production charm baryons in p--Pb collisions allow us to investigate if the hadronisation process is modified from pp to p--Pb collisions.

\section{Charm-hadron production}

\begin{figure}[t!]
    \centering
    \includegraphics[width=6cm]{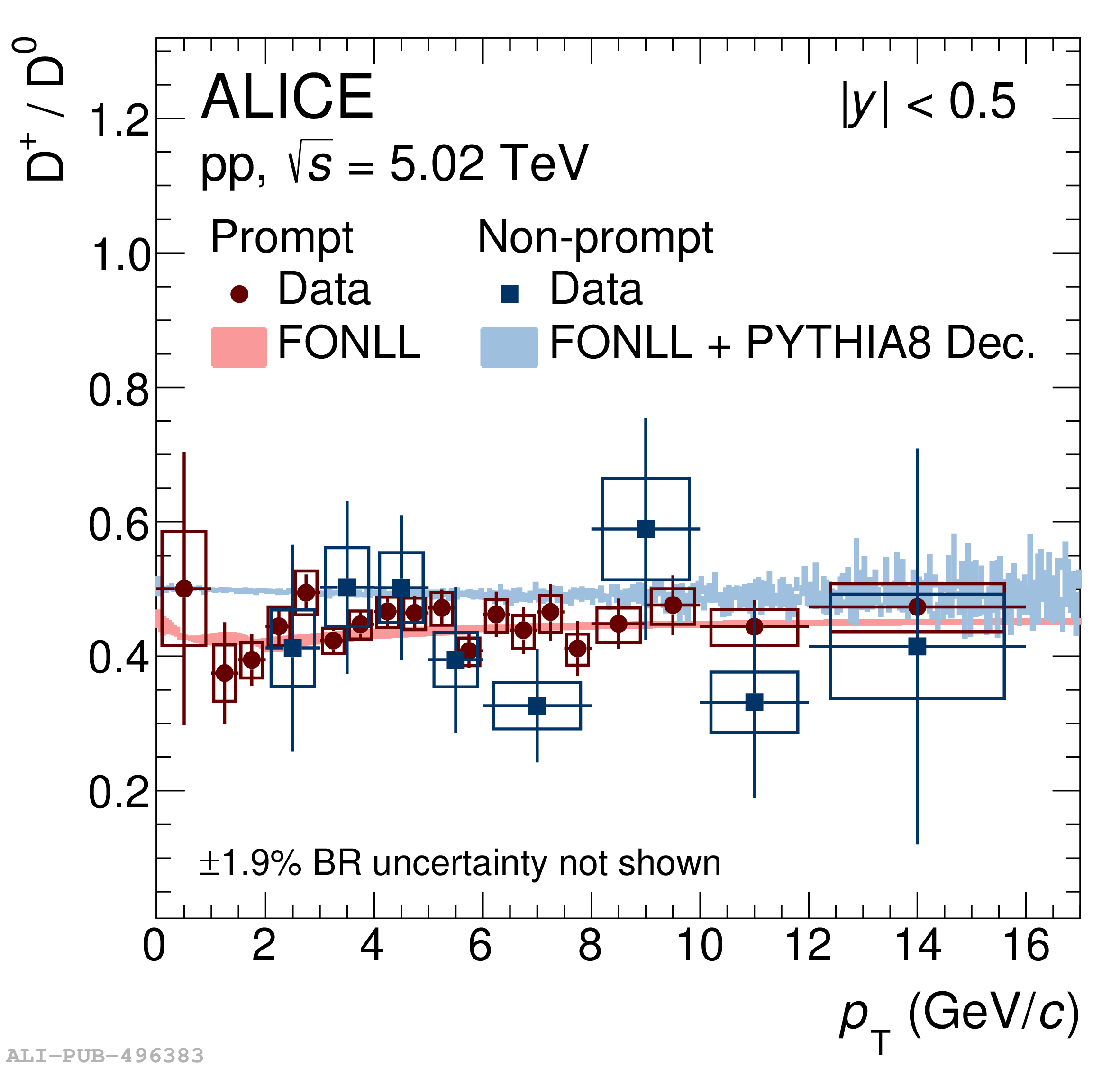}
    \includegraphics[width=6cm]{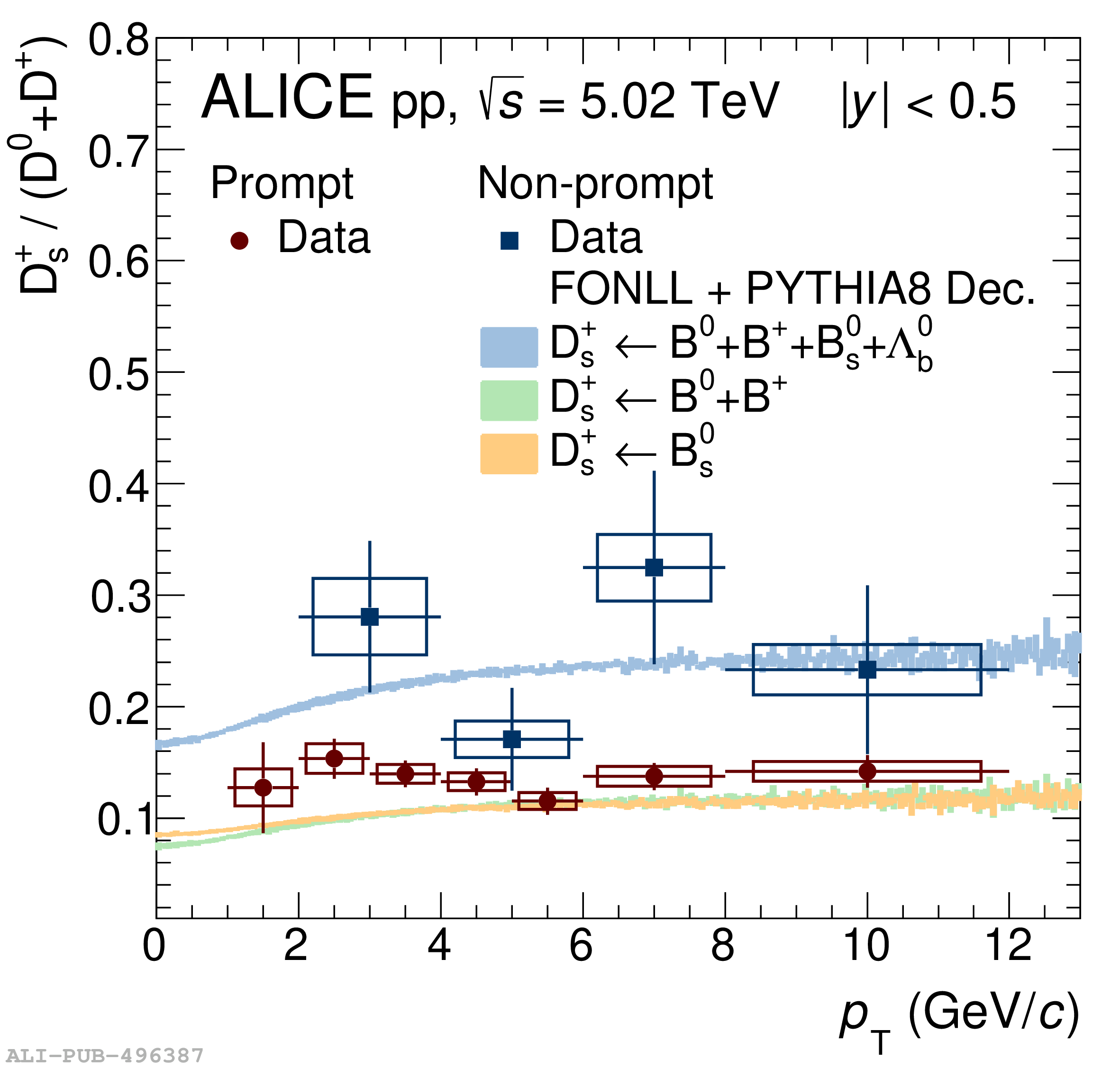}
    \caption{Left: ratios between the $p_{\rm T}$-differential production cross section of $\rm D^{+}$ and $\rm D^{0}$ mesons~\cite{ALICE:2021mgk}. Right: ratios between the $\rm D^{+}_{s}$-meson and the sum of the $\rm D^{0}$- and $\rm D^{+}$-meson production cross sections. The ratios are compared with prediction obtained with FONLL calculations~\cite{Cacciari:1998it,Cacciari:2001td} using PYTHIA 8~\cite{Sjostrand:2006za,Sjostrand:2014zea} for the $\rm H_{b} \rightarrow D + X$ decay kinematics. For the non-prompt $\rm D^{+}_{s}/(D^{0} + D^{+})$ ratio, the predictions for the $\rm D^{+}_{s}$ from $\rm B^{0}_{s}$ and from non-strange B-meson decays are also displayed separately.}
    \label{fig:D_meons}
\end{figure}

Recently, the ALICE Collaboration reported comprehensive measurements of the ground-state charm hadrons, the charm mesons $\rm D^{0}$, $\rm D^{+}$, $\rm D^{+}_{s}$, $\rm D^{*+}$\cite{ALICE:2021mgk} and the charm baryons $\Lambda_{\rm c}^{+}$, $\Sigma_{\rm c}^{0,++}$, $\Xi^{0,+}_{\rm c}$, $\Omega^{0}_{\rm c}$~\cite{ALICE:2022exq,ALICE:2021rzj,ALICE:2021bli,ALICE:2021psx,ALICE:2022cop}. The $p_{\rm T}$-differential cross sections of prompt and non-prompt (coming from B-hadron decays) D mesons~\cite{ALICE:2021mgk} have been compared with predictions from FONLL~\cite{Cacciari:1998it,Cacciari:2001td,Kniehl:2012ti} and GM-VFNS~\cite{Benzke:2017yjn,Kramer:2017gct,Bolzoni:2013vya} pQCD calculations. FONLL describes well both the results from prompt and non-prompt D mesons, while GM-VFNS is in good agreement with the prompt D mesons but underpredicts the non-prompt case. From these data, the ratios between the $p_{\rm T}$-differential production cross sections of $\rm D^{+}$ and $\rm D^{0}$ mesons and between the  $\rm D^{+}_{s}$ mesons and the sum of the $\rm D^{+}$ and $\rm D^{0}$ can be calculated, as shown in the left and right panels of Fig.~\ref{fig:D_meons}, respectively. Both the prompt and non-prompt meson-to-meson yield ratios are independent of the transverse momentum and are well described by pQCD calculations using fragmentation functions measured in $\rm e^{+}e^{-}$ and $\rm e^{-}p$ collisions~\cite{ALICE:2021mgk}. 

\begin{figure}[t!]
    \centering
    \includegraphics[width=6cm]{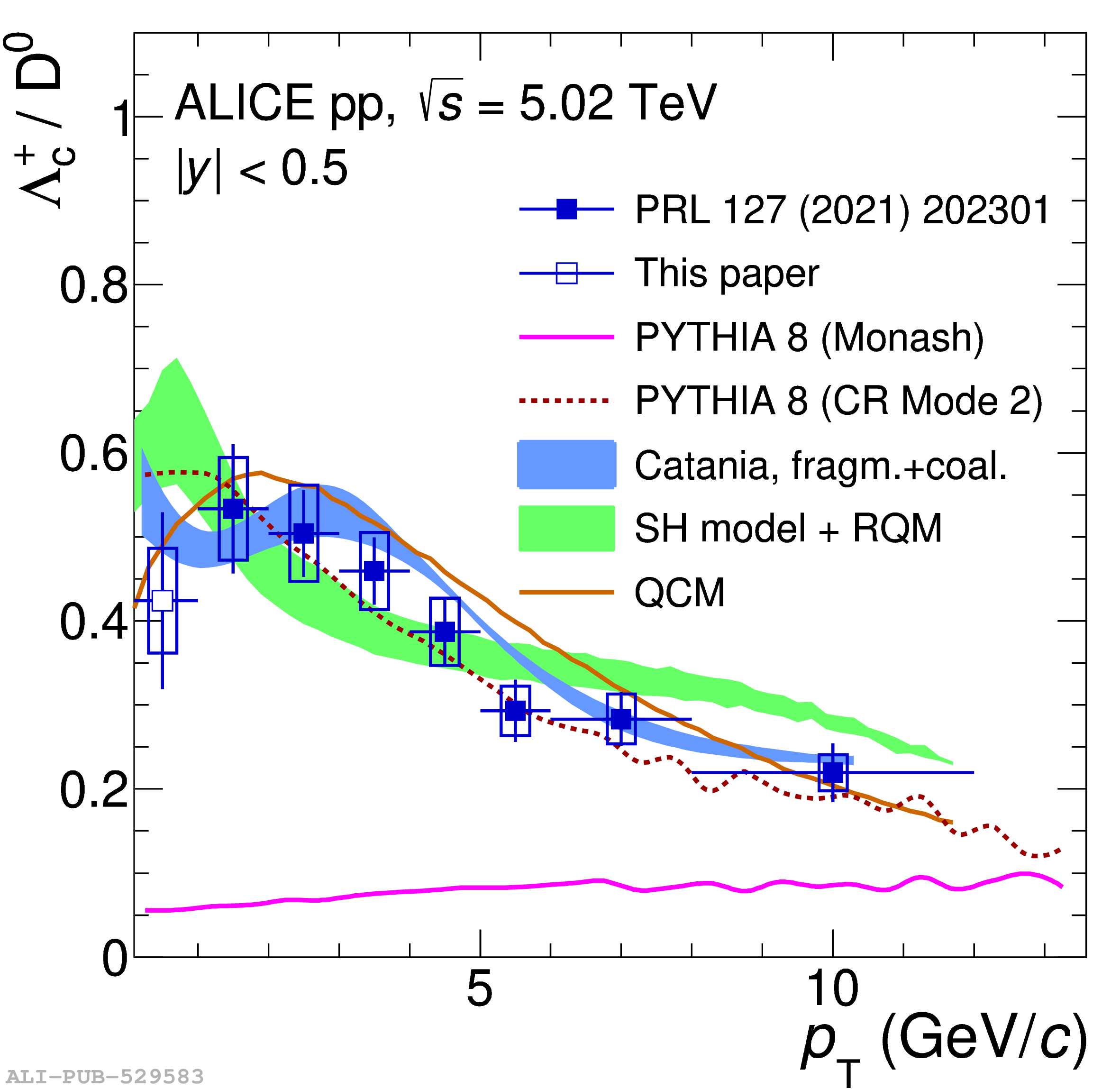}
    \includegraphics[width=6cm]{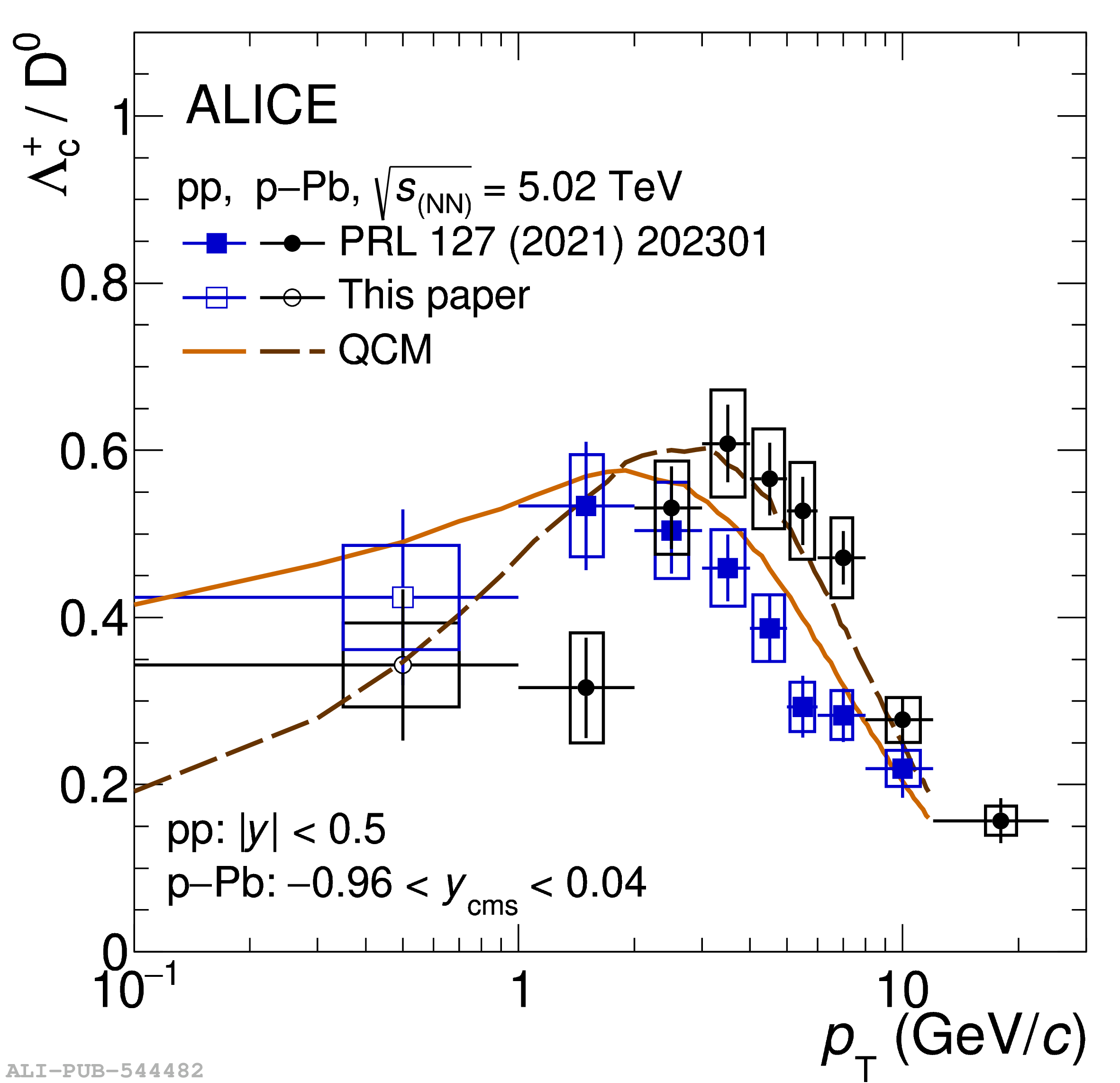}
    \caption{Left: the $\Lambda_{\rm c}^{+}/{\rm D^{0}}$ ratio as a function of $p_{\rm T}$ measured in pp collisions at $\sqrt{s} =$ 5.02 TeV compared with theoretical calculations (see text for details).  Right: prompt $\Lambda_{\rm c}^{+}/{\rm D^{0}}$ yield ratio in pp and p--Pb collisions as a function of $p_{\rm T}$, compared with QCM expectations~\cite{ALICE:2022exq}.}
    \label{fig:Lambda_D0}
\end{figure}

The $\Lambda_{\rm c}^{+}/{\rm D^{0}}$ yield ratio is used to further study the differences in the charm-quark hadronisation into mesons and baryons. The left panel of Fig.~\ref{fig:Lambda_D0} shows the $\Lambda_{\rm c}^{+}/{\rm D^{0}}$ yield ratios as a function of $p_{\rm T}$ down to $p_{\rm T} =$ 0~\cite{ALICE:2022exq}, which provides an important opportunity to study the charm hadronisation mechanism, as a significant fraction of the total charm-hadron yield is at $p_{\rm T} < 1$ GeV$/c$. The data are compared with different model predictions implementing different hadronisation processes.  PYTHIA 8 Monash, based on string fragmentation tuned to measurements in $\rm e^{+}e^{-}$ collisions, largely underestimates the data. However, the ratio is qualitatively described by model calculations including processes that enhance baryon production: 1) a tune of PYTHIA 8 with colour reconnection beyond the leading colour approximation, in which junction topologies increase the baryon production; 2) the Catania model implementing charm-quark hadronisation via both coalescence and fragmentation; 3) the Quark (re)-Combination Model (QCM), in which charm quarks convert into hadrons by recombining with light quarks; 4) the Statistical Hadronisation Model (SHM),  including feed-down to the ground-state charm baryon species from the decays of yet-unmeasured  resonant states predicted by the Relativistic Quark Model (RQM). The measurement of $p_{\rm T}$-differential $\Lambda_{\rm c}^{+}/{\rm D^{0}}$ production yield ratios in pp and p--Pb collisions~\cite{ALICE:2022exq} are shown in the right panel of Fig.~\ref{fig:Lambda_D0}. Within uncertainties, the $\Lambda_{\rm c}^{+}/{\rm D^{0}}$ ratios agree between the two collision systems, though the distributions peak in the region $1 < p_{\rm T} < 3$ GeV$/c$ in pp, and $3 < p_{\rm T} < 5$ GeV$/c$ in p--Pb collisions. The shifted peak from pp to p--Pb collisions could be due to the contribution of  collective expansion effects, e.g. radial flow. The results are compared with predictions from the QCM model, which describes the magnitude of the $\Lambda_{\rm c}^{+}/{\rm D^{0}}$ yield ratio well for $p_{\rm T} < $ 12 GeV$/c$ in both collision systems, and predicts a shifted peak towards higher $p_{\rm T}$, resulting from a hardening of the $\Lambda_{\rm c}^{+}$ spectrum in p--Pb collisions.

\section{Charm fragmentation fractions}

\begin{figure}
    \centering
    \includegraphics[width=6cm]{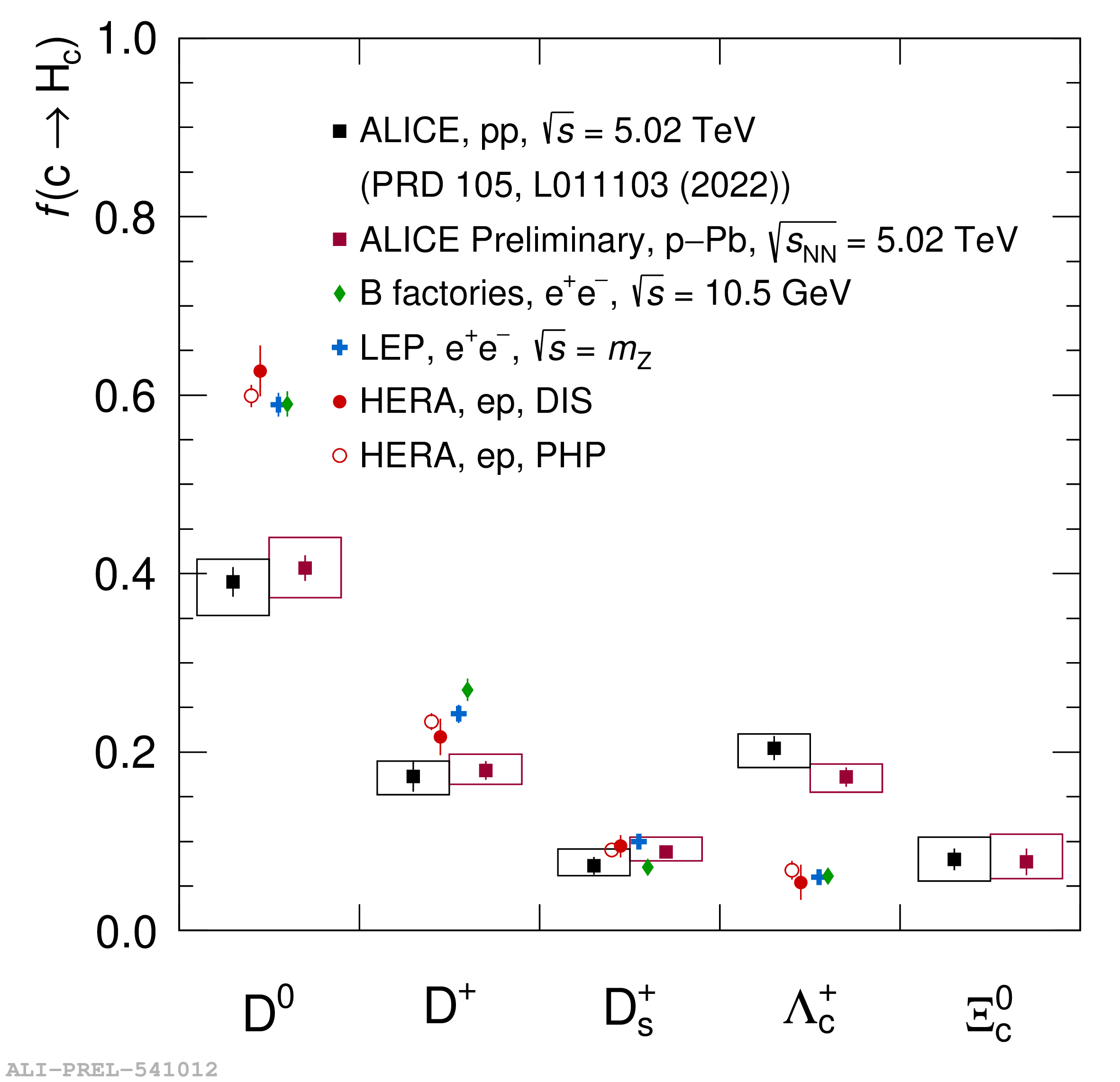}
    \includegraphics[width=6cm]{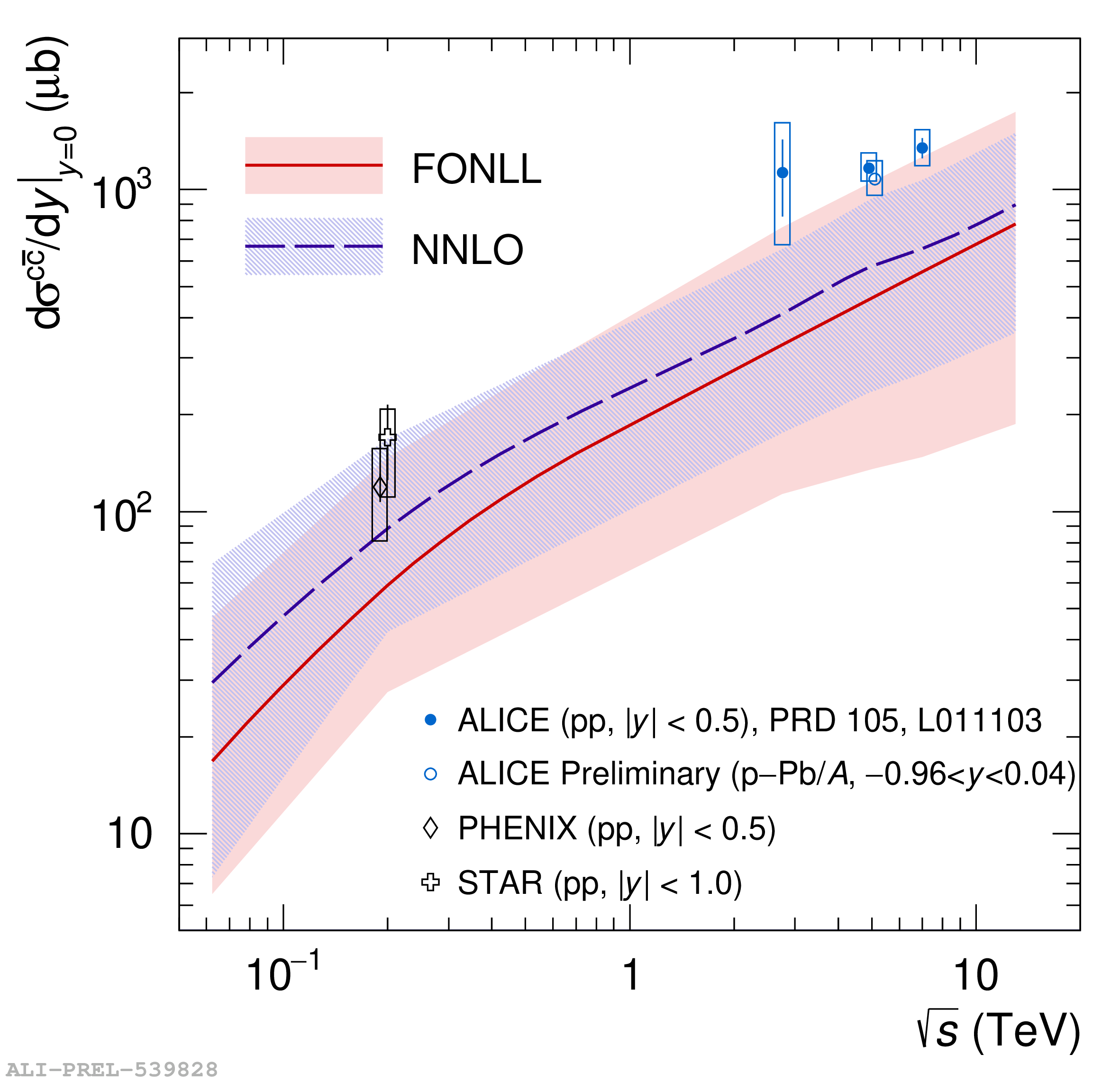}
    \caption{Left: charm-quark fragmentation fractions measured in pp and p–Pb collisions at $\sqrt{s_{\rm NN}}$ = 5.02 TeV compared with experimental measurements performed in $\rm e^{+}e^{-}$ collisions at LEP and B factories, and in $\rm e^{\pm}$p at HERA~\cite{Lisovyi:2015uqa,ALICE:2021dhb}. Right: total charm production cross section at midrapidity per unit of rapidity as a function of the collision energy at the LHC~\cite{ff_pp_ALICE,ffALICE276} and RHIC~\cite{STAR_ff} compared with FONLL~\cite{FONLL} and NNLO calculations~\cite{NNLO}.}
    \label{fig:ff}
\end{figure}

The charm fragmentation fractions $f(\rm c \rightarrow H_{c})$, shown in the left panel of Fig.~\ref{fig:ff}, represent the probabilities of a charm quark to hadronise into given charm-hadron species. The charm fragmentation fractions measured in pp and p--Pb collisions agree with each other. This suggests that the hadronisation mechanisms are not modified from pp to p--Pb collisions. In the right panel of Fig.~\ref{fig:ff}, the total $\rm c\bar{c}$ production cross section per unit of rapidity at midrapidity (${\rm d}\sigma^{\rm c\bar{c}}/{\rm d}y |_{|y| < 0.5}$) in pp~\cite{ALICE:2021dhb} and p--Pb collisions is shown as calculated by summing the $p_{\rm T}$-integrated cross sections of all measured ground-state charm hadrons ($\rm D^{0}$,  $\rm D^{+}$,  $\rm D^{+}_{s}$, $\Lambda^{+}_{\rm c}$, $\Xi^{0,+}_{\rm c}$ and their charge conjugates) in ALICE. The result in p--Pb collisions, scaled by the Pb mass number, is compatible with the one in pp collisions. The values are compared with FONLL~\cite{FONLL} and NNLO~\cite{NNLO} predictions as a function of the collision energy and lie on the upper edge of the theoretical pQCD calculations.

\section{Conclusions}

Recent measurements of charm-baryon production give stringent constraints to theoretical pQCD calculations. Charm hadronisation mechanisms in hadronic collisions need to be further investigated, in particular for the baryon sector. The charm fragmentation fractions are found to be different from those measured in  $\rm e^{+}e^{-}$ and $\rm e^{-}$p collisions, suggesting that the hadronisation mechanisms are not a universal process and depend on the collision system. More precise measurements that could shed further light into these topics are expected to be performed during Run 3 and 4 of the LHC, thanks to the ALICE detector upgrades~\cite{ALICE:2012dtf}.

\section{Acknowledgments}
This work is supported by the National Natural Science Foundation of China (No. 12275103, No.12061141008, and No.12105109).

\begin{adjustwidth}{-\extralength}{0cm}

\reftitle{References}

\bibstyle{Definitions/chicago2}
\bibliography{template}


\end{adjustwidth}
\end{document}